\documentstyle[12pt]{article}

\newcommand{\be}{\begin{equation}}
\newcommand{\ee}{\end{equation}}
\newcommand{\bea}{\begin{eqnarray}}
\newcommand{\eea}{\end{eqnarray}}
\newcommand{\sdeq}{Schwinger-Dyson equation}
\newcommand{\ddj}[1]{{\partial\over \partial J({#1})}}
\newcommand{\ddja}[1]{{\partial\over \partial J_0({#1})}}
\newcommand{\ddjb}[1]{{\partial\over \partial J_1({#1})}}
\newcommand{\ddjc}[1]{{\partial\over \partial J_2({#1})}}
\newcommand{\djj}[2]{{\partial^2\over \partial J({#1})\partial J({#2})}}
\newcommand{\djja}[2]{{\partial^2\over \partial J_0({#1})\partial J_0({#2})}}
\newcommand{\djjb}[2]{{\partial^2\over \partial J_0({#1})\partial J_1({#2})}}
\newcommand{\de}[1]{\delta_{{#1},0}}
\newcommand{\dd}[1]{{\delta\over\delta{#1}}}
\newcommand{\la}{\lambda}
\newcommand{\pl}{|+\rangle}
\newcommand{\hp}{{\cal H}(0)|+\rangle}

\begin{document}


\renewcommand{\thefootnote}{\fnsymbol{footnote}}

\begin{center}

\hfill KEK-TH-424\\
\hfill hep-th/9412120\\
\hfill December 1994\\
\vskip .5in

{\large \bf The Continuum Limit of the Schwinger-Dyson
            Equations of the One and Two Matrix Model
            with Finite Loop Length}
\vskip .8in

{\bf Michio Ikehara}\footnote{E-mail address: ikehara@theory.kek.jp}\\
\vskip
 .3in

{\em KEK theory group, Tsukuba, Ibaraki 305, Japan}\\
and\\
{\em Department of Physics, University of Tokyo, Bunkyo-ku, Tokyo 113, Japan}\\

\vskip .5in

\end{center}


\begin{abstract}

We take the continuum limit of the \sdeq s of the one and two
matrix model without expanding them in the length of the loop.
The resulting equations agree with those proposed for
string field theory in the temporal gauge.
We find that the loop operators are required to mix
in the two matrix model case
and determine the non-constant tadpole terms.

\end{abstract}

\renewcommand{\thefootnote}{\arabic{footnote}}
\setcounter{footnote}{0}

\newpage

The Virasoro and the $W_3$\cite{fkn} constraints were
obtained by taking the continuum limit of the one and the
two\cite{gn} matrix model, respectively.
Motivated by the transfer matrix formalism\cite{kkmw}
string field theory in the temporal gauge was
proposed for $c=0$\cite{IK} and $c=1-6/m(m+1)$\cite{IK2}\cite{iikmns}.
In \cite{IK} the corresponding Virasoro constraints were derived from
the \sdeq\ reproducing the results of the matrix model.
In \cite{iikmns} the \sdeq s for $c=1-6/m(m+1)$ were proposed in
the continuum theory, allowing for the derivation of the $W_m$ constraints.
It was checked that the continuum limit of the dynamical triangulation
becomes $c=0$ string field theory\cite{W}.
In this paper we take the continuum limit of the one and
two matrix model step by step.
We directly derive the \sdeq s in
\cite{IK} and \cite{iikmns}.
In this approach we do not expand correlation functions
in terms of the length of the loop.

We begin with the one matrix $\phi^3$ model.
The expectation value of some quantity in this model is:
\be \langle \cdots \rangle = {\int d\phi \cdots \exp \left( -N \mbox{tr}\left(
{1\over 2} \phi^{2}-{\lambda\over 3}\phi^{3}\right)\right)\over
\int d\phi \exp \left( -N \mbox{tr}\left(
{1\over 2} \phi^{2}-{\lambda\over 3}\phi^{3}\right)\right)}
, \label{mm}\ee
where $\phi$ is $N\times N$ hermitian matrix.
The dual graph of this model consists of triangles whose sides are attached
to another side of some triangle.
We assume that these are equilateral triangles.
We define the partition function of the loop amplitudes in the matrix model as
\be Z_m(J)=\left\langle \exp\left(\sum_{n} J(n)W(n)\right)\right\rangle, \ee
where $W(n)$ is the loop operator in the matrix model defined as
\be W(n)=\left\{ \begin{array}{ll}
    {1\over N}\mbox{tr}\phi^{n}& (n\geq 0)\\
                0 & (n \leq -1).
         \end{array}\right.\label{wn}\ee

The \sdeq\  of the matrix model is obtained by taking the total derivative
with respect to $\phi$ inside the integration.
\be \left\langle {1\over N^2}\mbox{tr}\left({\partial\over{\partial \phi}}
\phi^{n-1}\right)\exp\left(\sum_{m} J(m)W(m)\right)\right\rangle=0.\ee
Here $n\geq 1$ and the differentiation with respect to $\phi$ operates
not only to the inside of the parenthesis but also to the outside,
including the action in eq.(\ref{mm}).
In terms of $Z_m(J)$, this can be written as
\bea \lefteqn{}&&\left[\sum_{m}\djj{m}{n-2-m}+{\theta(n)\over N^{2}}
\sum_{m=1}^\infty mJ(m)\ddj{n+m-2}\right.\nonumber\\
&&\left. \mbox{}-\ddj{n}+\la\ddj{n+1}+\de{n}\left(1-
\la{\partial\over\partial J(1)}
\right)-\la\de{n+1}\right]Z_m(J)=0,\label{sd0}\eea
where
\be \theta(n)=\left\{\begin{array}{ll}
                        1&(n\geq1)\\
                        0&(n\leq0).\end{array}\right. \ee
We added $\theta(n)$ and the Kronecker $\delta$'s
in order that eq.(\ref{sd0}) will hold for all integer $n$.
We also replaced $\partial/\partial J(0)$ by 1, which is evident
in eq.(\ref{wn}).
Since the term $\de{n}\partial/\partial J(1)$ complicates the
following argument, we choose to drop this term by multiplying
eq.(\ref{sd0}) by $n$.

In the continuum limit, the $\delta$-function and its derivatives
in terms of the loop length will appear in $W(n)$.
To obtain the continuum loop operator we have to subtract
these terms from $W(n)$.
We therefore employ the following partition function $Z_c(J)$:
\bea Z_m(J)&=&\exp\left(\sum_n J(n)c(n)\right)Z_c(J),\nonumber\\
Z_{c}(J)&=&\left\langle\exp\left(\sum_{n}J(n)(W(n)-c(n))\right)\right
\rangle,\label{z0}\eea
where $c(n)$ consists of Kronecker $\delta$'s which become in the
continuum limit the $\delta$-functions of the loop length.
Using $Z_{c}$, eq.(\ref{sd0}) mutiplied by $n$ can be rewritten as
\bea \lefteqn{} & &\left[n\sum_{m}\djj{m}{n-2-m}+
{n\theta(n)\over N^{2}}\sum_{m=1}^\infty mJ(m)\ddj{n+m-2}\right.\nonumber\\
& &\mbox{}-n\ddj{n}+n\la\ddj{n+1}+2n\sum_{m}c(m)\ddj{n-2-m}\nonumber\\
& &\mbox{}+{n\theta(n)\over N^{2}}\sum_{m=1}^\infty mJ(m)c(n+m-2)
+n\sum_{m}c(m)c(n-2-m)\nonumber\\
& &\left.\mbox{}-nc(n)+n\la c(n+1)+\la\de{n+1}\right]
Z_{c}(J)=0.\label{sd00}\eea

We want to take the continuum limit of this \sdeq.
We first define the loop length represented by the loop operator
$W(n)$ as $n$ times $a$, the length of the sides of the equilateral triangles.
\be l=na.\ee
In the continuum limit, $a\rightarrow 0$ with $l$ kept finite,
we can replace the sum over $n$ by the
integral over the loop length, and Kronecker $\delta$'s by $\delta$-functions:
\be \sum_n\rightarrow{1\over a}\int dl, \ee
\be \de{n+k}\rightarrow a\delta(l+ka).\ee
Next we consider what happens to the loop operator $W(n)$
in the continuum limit.
The dimension of the continuum disk amplitude is $L^{-5/2}$\cite{KPZDDK}
where $L$ is the dimension of the loop length.
We therefore assume that the loop operator $W(n)-c(n)$ scales
as $a^{5/2}$.
\be y_c^n(W(n)-c(n))\sim a^{5\over 2}w(l).\ee
Here, $w(l)$ is the continuum loop operator whose dimension is $L^{-5/2}$.
The symbol $\sim$ means that the right hand side is the leading
term in the limit $a\rightarrow 0$.
The factor $y_c^n$ is introduced so that the Laplace transformation
of the loop operator becomes:
\be \sum_n y^n(W(n)-c(n))\sim a^{3\over 2}\int_0^\infty
dlw(l)e^{-\zeta l},\ee
\be y=y_c e^{-a\zeta},\ee
where $y_c$ is the critical value of $y$ where we can take the double
scaling limit.
We introduce the string coupling constant $g$ as follows\cite{DSL}.
\be {1\over N^2}=a^5g.\label{scc}\ee
This $g$ counts the genus of the surface in the genus expansion.
The cosmological constant $t$ in the continuum limit is defined as
\be \la=\la_ce^{-a^2t},\label{cc}\ee
where $\la_c$ is the critical value of $\la$.
In order to obtain the continuum partition function,
we define
\be y_c^{-n}J(n)=a^{-{3\over 2}}j(l).\ee
We then find that
\be y_c^n\ddj{n}\sim a^{5\over 2}\dd{j(l)}.\ee
Using $j(l)$ and $w(l)$, the partition function $Z_c$ in eq.(\ref{z0})
becomes:
\be Z_c(J)\sim \left\langle \exp\left(\int_0^\infty dlj(l)w(l)\right)
\right\rangle\equiv Z[j],\ee
where $\langle\cdots\rangle$ now represents the continuum expectation
value.

Now we go back to eq.(\ref{sd00}). We multiply this equation by
$y_c^{n-2}$ and express all the quantities by the continuum ones.
Let us examine the order of each term.
The first and the second terms are of order $a^3$.
The linear terms of $\partial/\partial J$ and $J$ are of order $a^{3/2}$.
The terms involving only $c(n)$ and Kronecker $\delta$
turn to $\delta$-functions which can be expanded in the positive integer
powers of $a$.
We choose the critical values, $y_c$ and $\la_c$, so that
the terms having lower orders of $a$ than the resulting equation vanish.

Suppose that we choose a certain $c(n)$ and that the linear
terms of $\partial/\partial J$ remain.
These terms and the linear term of $J$, then, are the leading order terms.
The resulting equation, however, implies that the disk amplitude vanishes.
These terms therefore should vanish.
We can thus obtain $c(n)$ from the requirement that the linear terms of
$\partial/\partial J$ in eq.(\ref{sd00}) vanish. That is,
\be c(n)=-{\la\over 2}\de{n+3}+{1\over 2}\de{n+2}.\ee
The linear term of $J$ in eq.(\ref{sd00}) also vanishes
when $c(n)$ is in this form.

Thus the remaining terms are the first two terms and $\delta$-functions
in eq.(\ref{sd00}).
The first two terms have to be the leading order in the continuum limit,
so the terms of order $a$ and $a^2$ from the $\delta$-function
should vanish.
We solve the equations given by setting these terms zero.
We find
\bea y_c&=&\sqrt{{2\sqrt{3}-3}\over 6},\\
     \la_c&=&{1\over{2\cdot 3^{3\over 4}}},\eea
which agree with the previous result\cite{BIPZ}.
We thus obtain the continuum limit of the \sdeq\ from the $a^3$ term
in eq.(\ref{sd00}).
This can be expressed in a simple form by using the symbols
$*$ and $\triangleleft$ defined as
\be (f*h)(l)=\int_0^l dl'f(l')h(l-l'),\ee
\be (f\triangleleft h)(l)=\int_0^\infty dl'f(l')h(l+l').\ee
We further rescale the loop operator, the string coupling and the cosmological
constant for convenience.
\be w(l)\rightarrow {(1+\sqrt{3})^{5\over 2}\over 2}w(l),\ee
\be j(l)\rightarrow {2\over (1+\sqrt{3})^{5\over 2}}j(l),\ee
\be g\rightarrow \left({(1+\sqrt{3})^{5\over 2}\over 2}\right)^2g,\ee
\be t\rightarrow \left({3+\sqrt{3}}\over 4\right)^2t.\ee
Using these notations the continuum limit of the \sdeq\ is
\be \left[l\left(\dd{j}*+g(lj)\triangleleft \right)\dd{j}(l)+\rho(l)\right]
Z[j]=0,\label{csd0}\ee
where $\rho$ is the tadpole,
\be \rho(l)=3\delta^{(2)}(l)-{3\over 4}t\delta(l).\ee
This \sdeq\ agrees with that of the string field theory for
$c=0$\cite{IK}.
Actually the Heaviside function appears in eq.(\ref{csd0})
from the continuum limit of $\theta(n)$.
We dropped it from eq.(\ref{csd0}) by demanding that $l\geq 0$.

Following the procedure in \cite{IK} we can transform
eq.(\ref{csd0}) into the Virasoro constraints.
We here rewrite eq.(\ref{csd0}) into a form in which
the first term does not depend on the factor of $l$, that is,
\be \left[\left(\dd{j}*+(lj)\triangleleft \right)\dd{j}(l)
-\delta^{(3)}(l)+{3\over 4}t\delta^{(1)}(l)\right]
Z[j]+\delta(l)R[j]=0.\label{csd00}\ee
Here we have put $g=1$ for simplicity.
This kind of form will be necessary in the $c=1/2$ case.
In order to get the Virasoro constraints we subtract from $\log Z[j]$
the singular terms which cannot be expanded in
positive power of the loop length.
\be Z[j]=Z_{sin}[j]Z_{reg}[j],\ee
\be Z_{sin}[j]=\exp\left(\int_0^\infty dlj(l) w_s(l)
   +{1\over 2}\int_0^\infty dl\int_0^\infty dl'j(l)j(l')c_s(l,l')\right),\ee
\be w_s(l)={1\over \Gamma(-{3\over 2})}l^{-{5\over 2}}
           -{3t\over 8\Gamma({1\over 2})}l^{-{1\over 2}},\ee
\be c_s(l,l')={1\over 2\pi}{\sqrt{ll'}\over l+l'}.\ee
The regular loop amplitudes contained in $Z_{reg}[j]$ can be expanded
with respect to the loop lengths in powers of
a positive odd integer divided by 2.
We therefore can use the following variables instead of $j(l)$.
\be j_r=\int_0^\infty dlj(l){l^r\over \Gamma(1+r)},\label{jr}\ee
\be \dd{j(l)}Z_{reg}[j]=D(l)Z_{reg}[j],\ee
\be D(l)=\sum_r {l^r\over \Gamma(1+r)}{\partial\over\partial
j_r},\label{ddd}\ee
where
\be r\equiv {1\over 2}\pmod{1},\ee
\be r>0.\ee
In this notation eq.(\ref{csd00}) becomes
\be \left\{\left[\mbox{\Large :}\left(D(l)+{K(l)\over 2}+w_s(l)\right)^2
\mbox{\Large :}+{1\over 16}l\right]_{\geq 0}
    +\delta(l)\left(2{\partial\over\partial j_{1\over 2}}
    +{1\over 2}j_{-{1\over 2}}\right)
    \right\}Z_{reg}[j]+\delta(l){R[j]\over Z_{sin}[j]}=0,\label{vir}\ee
where
\be K(l)=\sum_r {l^{-r}\over \Gamma(1-r)}rj_r,\label{kkk}\ee
:: means the normal ordering of the differentiation,
$[\cdots]_{\geq 0}$ means the terms of $\cdots$ with the non-negative
integer power of $l$,
and $j_{-1/2}$ is defined as in eq.(\ref{jr}).
In eq.(\ref{vir}) the coefficients of the $l$ expansion
lead to the Virasoro constraints.
Here $w_s(l)$ can be viewed as the background of $j_r$
contained in $K(l)$.
The background of $j_{1\over 2}$ and $j_{5\over 2}$ is seen to be
$-3t/2$ and $4/5$, respectively\cite{fkn}.
The coefficient of the $\delta$-function in eq.(\ref{vir})
leads to
\be R[j]=-Z_{sin}[j]\left(2{\partial\over\partial j_{1\over 2}}
    +{1\over 2}j_{-{1\over 2}}
    \right)Z_{reg}[j].\ee
Namely $R[j]$ cannot be $Z[j]$ multiplied by a constant.
We can understand this by noticing the fact that
$R[j]$ corresponds to $1-\la\partial/\partial J(1)$ in eq.(\ref{sd0}).
Indeed we can verify that the disk amplitude obtained from eq.(\ref{csd0})
satisfies eq.(\ref{csd00}) with $R[j]$ given above.

We now go on to the two matrix model.
In order to obtain the continuum \sdeq s proposed in \cite{iikmns}
to derive the $W_3$ constraints,
we employ the loops with (i) only the same matrix,
(ii) a different matrix inserted and (iii) two different matrices inserted
next to each other.
We therefore define the partition function as
\be Z_m(J_0,J_1,J_2)=\left\langle\exp\left[\sum_{k=0}^2\sum_nJ_k(n)W_k(n)
\right]\right\rangle,\ee
where
\be \langle\cdots\rangle={\int d\phi_1 d\phi_2 \cdots \exp\left[
    -N\mbox{tr}\left\{\sum_{i=1}^2\left({1\over 2}\phi_i^2-{\la\over 3}
     \phi_i^3\right)-\mu \phi_1\phi_2\right\}\right]\over
     \int d\phi_1 d\phi_2 \exp\left[
    -N\mbox{tr}\left\{\sum_{i=1}^2\left({1\over 2}\phi_i^2-{\la\over 3}
     \phi_i^3\right)-\mu \phi_1\phi_2\right\}\right]},\ee
\be W_0(n)={1\over N}\mbox{tr}\phi_1^n,\ee
\be W_1(n)={1\over N}\mbox{tr}\phi_1^{n-1}\phi_2,\ee
\be W_2(n)={1\over N}\mbox{tr}\phi_1^{n-2}\phi_2^{\, 2},\ee
\be W_0(n)=W_1(n+1)=W_2(n+2)=0 \mbox{ for }n\leq -1.\ee
We consider the following three \sdeq s\cite{iikmns}.
\be \left\langle {1\over N^2}\mbox{tr}\left({\partial\over{\partial \phi_1}}
\phi_1^{n-1}\right)\exp\left(\sum_k\sum_{m} J_k(m)W_k(m)\right)\right\rangle
=0\; (n\geq 1),\ee
\be \left\langle {1\over N^2}\mbox{tr}\left({\partial\over{\partial \phi_1}}
\phi_1^{n-2}\phi_2\right)\exp\left(\sum_k\sum_{m} J_k(m)W_k(m)\right)\right
\rangle=0\; (n\geq 2),\ee
\be \left\langle {1\over N^2}\mbox{tr}\left({\partial\over{\partial \phi_2}}
\phi_1^{n-1}\right)\exp\left(\sum_k\sum_{m} J_k(m)W_k(m)\right)\right\rangle
=0\; (n\geq 1).\ee
We can write these equations in terms of the partition function as
\bea \lefteqn{}&&\left[\sum_{m}\djja{m}{n-2-m}+{\theta(n)\over N^{2}}
\sum_{m=1}^\infty mJ_0(m)\ddja{n+m-2}\right.\nonumber\\
&&\mbox{}-\ddja{n}+\la\ddja{n+1}+\mu\ddjb{n}
+\de{n}\left(1-\la{\partial\over\partial J_0(1)}\right)\nonumber\\
&&\mbox{}\left.-\la\de{n+1}\right]
Z_m(J_0,J_1,J_2)|_{J_1=J_2=0}=0,\label{sd1}\eea
\bea \lefteqn{}&&\left[\sum_{m}\djjb{m}{n-2-m}+{\theta(n-1)\over N^{2}}
\sum_{m=1}^\infty mJ_0(m)\ddjb{n+m-2}\right.\nonumber\\
&&\mbox{}-\ddjb{n}+\la\ddjb{n+1}+\mu\ddjc{n}-\de{n}\la{\partial\over\partial
J_1(1)}
\nonumber\\
&&\mbox{}\left.+\de{n-1}\left({\partial\over\partial J_1(1)}-\la
{\partial\over\partial J_1(2)}\right)\right]
Z_m(J_0,J_1,J_2)|_{J_1=J_2=0}=0,\label{sd2}\eea
\be \left[-\ddjb{n}+\la\ddjc{n+1}+\mu\ddja{n}-\mu\de{n}\right]
Z_m(J_0,J_1,J_2)|_{J_1=J_2=0}=0.\label{sd3}\ee
These equations now hold for all integer $n$.
We can solve eq.(\ref{sd3}) for $\partial/\partial J_2$ and
substitute it into eq.(\ref{sd2}):
\bea \lefteqn{}&&\left[\sum_{m}\djjb{m}{n-2-m}+{\theta(n-1)\over N^{2}}
\sum_{m=1}^\infty mJ(m)\ddjb{n+m-2}\right.\nonumber\\
&&\mbox{}-\ddjb{n}+\la\ddjb{n+1}+{\mu\over \la}\ddjb{n-1}
-{\mu^2\over \la}\ddja{n-1}-\de{n}\la {\partial\over\partial J_1(1)}
\nonumber\\
&&\mbox{}\left.
+\de{n-1}\left({\partial\over\partial J_1(1)}-\la{\partial\over\partial J_1(2)}
+{\mu^2\over \la}
\right)\right]Z_m(J_0,J_1,J_2)|_{J_1=J_2=0}=0.\label{sd23}\eea
This equation and eq.(\ref{sd1}) now contain no $J_2$.
We thus omit $J_2$ in the following.

Unlike the case of the one matrix model, the problem of operator mixing
appears here.
In order to see what happens, we investigate the scaling dimensions
of loop operators.
The loop operator $W_0(n)$ minus $c_0(n)$ scales as\cite{mss}
\be y_c^n(W_0(n)-c_0(n))\sim a^{7\over 3}w(l,\pl).\ee
Here $w(l,\pl)$ is the continuum loop operator for the loop whose
length is $l$ and whose spin state is $\pl$, that is, all up.
At first sight $W_1(n)$ seems to become, in the continuum limit,
the operator for the loop whose spin is up on almost all
the points except for an infinitesimal down region.
In \cite{iikmns} the dimension of such a loop was found to be $L^{-11/3}$.
We first expect that
\be y_c^n(W_1(n)-c_1(n))\stackrel{?}{\sim} a^{11\over 3}w(l,\hp),
\label{w1}\ee
where $\hp$ is the state whose spin is flipped at the origin.
Similar to the case of the one matrix model, the linear terms of
$\partial/\partial J_0$ in eqs.(\ref{sd1}) and (\ref{sd23})
should vanish after the subtraction of $c$'s.
We can solve $c_0$ and $c_1$ from these conditions.
However we find that the linear terms of $\partial/\partial J_1$
in eq.(\ref{sd23}) does not vanish for this $c_0$.
We would therefore arrive at the conclusion that $w(l,\hp)=0$ if we
adopted eq.(\ref{w1}).

In order to obtain the correct result we first subtract $c_0(n)$ from
$W_0(n)$ only,
eq.(\ref{sd1}) then becomes
\bea \lefteqn{}&&\left[\sum_{m}\djja{m}{n-2-m}+{\theta(n)\over N^{2}}
\sum_{m=1}^\infty mJ_0(m)\ddja{n+m-2}\right.\nonumber\\
&&\mbox{}-\ddja{n}+\la\ddja{n+1}+\mu\ddjb{n}+2\sum_m c_0(m)\ddja{n-2-m}
\nonumber\\
&&\mbox{}+{\theta(n)\over N^{2}}\sum_{m=1}^\infty mJ(m)c_0(n+m-2)
+\de{n}\left(1-\la{\partial\over\partial J_0(1)}\right)-\la\de{n+1}\nonumber\\
&&\left.-c_0(n)+\la c_0(n+1)+\sum_m c_0(m)c_0(n-2-m)
\right]Z_c'(J_0,J_1)|_{J_1=0}=0.
\label{sd11}\eea
We introduce the string coupling constant $g$ as\cite{DSL}
\be {1\over N^2}=a^{14\over 3}g.\label{ngg}\ee
The first two terms in eq.(\ref{sd11}) are of order $a^{11/3}$.
They are the same order as $a^{11/3}w(l,\hp)$.
We therefore demand that the linear terms of $\partial/\partial J_0$
and $\partial/\partial J_1$ in eq.(\ref{sd11}) together create $w(l,\hp)$.
We define the corresponding loop operator as
\be V_1(n)={1\over\mu}\left[\mu W_1(n)-W_0(n)+\la W_0(n+1)+2\sum_m
c_0(m)W_0(n-2-m)\right],\ee
\be y_c^n(V_1(n)-c_1(n))\sim a^{11\over 3}w(l,\hp).\ee
This means that we have to mix $W_1$ with $W_0$ to obtain
$w(l,\hp)$ in the continuum limit.
Changing variables for $Z$ and subtracting $c$'s
we get the new partition function.
\be Z_c(J_0,J_1)=\left\langle\exp\sum_n\left\{J_0(n)(W_0(n)-c_0(n))
+J_1(n)(V_1(n)-c_1(n))\right\}\right\rangle.\ee
We now can express the linear terms of $\partial/\partial J_0$
and $\partial/\partial J_1$ in
eq.(\ref{sd11}) as $\partial/\partial J_1$ in terms of this $Z_c(J_0,J_1)$.

Rewriting eq.(\ref{sd23}) using $Z_c$, eliminating the quadratic terms of
$\partial/\partial J_0$ by eq.(\ref{sd1}) rewritten by $Z_c$ and
demanding that the linear terms of $\partial/\partial J_0$ and
$\partial/\partial J_1$ vanish, we find
\be c_0(n)=-{2\la\over 3}\de{n+3}+{2\over 3}\de{n+2}-{\mu\over
3\la}\de{n+1},\ee
\be c_1(n)={\la^2\over 9\mu}\de{n+4}-{2\la\over 9\mu}\de{n+3}
     +{\mu +1\over 9\mu}\de{n+2}-{1-9\mu^2\over 9\la}\de{n+1}
    -{2\mu\over 9\la^2}\de{n}.\ee
The linear terms of $J_0$ in eqs.(\ref{sd1}) and (\ref{sd23}) rewritten
by using $Z_c$ vanish with these $c$'s.
The \sdeq s therefore become
\bea \lefteqn{}&&\left[\sum_{m}\djja{m}{n-2-m}+{\theta(n)\over N^{2}}
\sum_{m=1}^\infty mJ_0(m)\ddja{n+m-2}+\mu\ddjb{n}\right.\nonumber\\
&&\mbox{}-{\la^2\over 3}\de{n+4}+{2\la\over 3}\de{n+3}-{\mu+1\over 3}\de{n+2}
+{\mu(1+3\mu^2)-3\la^2\over 3\la}\de{n+1}\nonumber\\
&&\mbox{}\left.+\de{n}\left(1-\la{\partial\over\partial J_0(1)}
-{\mu^2\over 3\la^2}\right)
\right]Z_c(J_0,J_1)|_{J_1=0}=0,
\label{sd111}\eea
\bea \lefteqn{}&&\left[\sum_{m}\djjb{m}{n-2-m}+{\theta(n-1)\over N^{2}}
\sum_{m=1}^\infty mJ_0(m)\ddjb{n+m-2}\right.\nonumber\\
&&\mbox{}+{1\over N^{2}}\sum_{m=1}^\infty mJ_0(m)\left({\de{n-1}-\la\de{n}\over
3\mu}\ddja{m-1}-{\la\de{n-1}\over 3\mu}\ddja{m}\right)
\nonumber\\
&&\mbox{}+{2\la^3\over 27\mu}\de{n+5}
-{2\la^2\over 9\mu}\de{n+4}+{\la(\mu+2)\over 9\mu}\de{n+3}
+{1\over 3\mu}\left(2\mu^3-{2\mu\over 3}+\la^2-{2\over 9}\right)\de{n+2}
\nonumber\\
&&\mbox{}+\de{n+1}\left({1-\mu-6\mu^2\over 9\la}-{\la\over 3\mu}
\left(2-\la\ddja{1}\right)\right)\nonumber\\
&&\mbox{}
+\de{n}\left({2\over 3}+{1\over 3\mu}+{\mu(1+3\mu^2)\over 9\la^2}
-\la\ddjb{1}-{\la\over 3\mu}\ddja{1}\right)\nonumber\\
&&\mbox{}\left.
+\de{n-1}\left({\mu^2\over\la}\left(1-{2\over 27\la^2}\right)-{2\over 3\la}
\left(1-\la\ddja{1}\right)+\ddjb{1}-\la\ddjb{2}\right)
\right]\nonumber\\
&& Z_c(J_0,J_1)|_{J_1=0}=0.
\label{sd222}\eea
The terms involving $\partial/\partial J_0(1)$,
$\partial/\partial J_1(1)$ and $\partial/\partial J_1(2)$
in eqs.(\ref{sd111}) and (\ref{sd222})
complicate the following argument.
In order to drop these terms
we multiply eqs.(\ref{sd111}) and (\ref{sd222}) by $ny_c^{n-2}$ and
$n(n^2-1)y_c^{n-2}$, respectively.
The terms like
\be -{\la\de{n-1}\over 3\mu N^2}\sum_m mJ_0(m)\ddja{m}\ee
in eq.(\ref{sd222}) also vanish by this multiplication.

In order to take the continuum limit we substitute eqs.(\ref{cc})
and (\ref{ngg}) into eqs.(\ref{sd111}) and (\ref{sd222}).
We define
\be y_c^{-n}J_0(n)=a^{-{4\over 3}}j_0(l),\ee
\be y_c^{-n}J_1(n)=a^{-{8\over 3}}j_1(l).\ee
It follows that
\be y_c^n\ddja{n}\sim a^{7\over 3}\dd{j_0(l)},\ee
\be y_c^n\ddjb{n}\sim a^{11\over 3}\dd{j_1(l)}.\ee
We use the continuum quantities and find that the first three terms
in eq.(\ref{sd111}) turn to
the order of $a^{8/3}$ by the multiplication of $n$.
The first two terms of eq.(\ref{sd222})
turn to the order of $a^2$ by the multiplication of $n(n^2-1)$.
These terms have to be the leading order in the continuum limit.
We thus obtain three conditions: the terms of $a$ and $a^2$
in eq.(\ref{sd111}) and
the term of $a$ in eq.(\ref{sd222}) must vanish.
We find the critical values from these conditions.
\be \mu={2\sqrt{7}-1\over 27},\ee
\be \la_c={1\over 243}\sqrt{30(62\sqrt{7}-85)},\ee
\be y_c={1\over 27}\sqrt{30(26\sqrt{7}-67)}.\ee
The above values agree with the previous result\cite{bk}.
We rescale $w(l,\pl)$, $w(l,\hp)$, $j_0(l)$, $j_1(l)$, $g$ and $t$
for convenience.
We find from the $a^{8/3}$ and $a^2$ terms of eqs.(\ref{sd111})
and (\ref{sd222}), respectively,
that the continuum \sdeq s are
\be l\left[\left(\dd{j_0}*+g(lj_0)\triangleleft\right)\dd{j_0}(l)
+\dd{j_1(l)}\right]Z[j_0,j_1]|_{j_1=0}=0,\label{csd1}\ee
\be \left[l^3\left(\dd{j_0}*+g(lj_0)\triangleleft\right)\dd{j_1}(l)
-384\delta^{(1)}(l)\right]Z[j_0,j_1]|_{j_1=0}=0.\label{csd2}\ee
These equations agree with those which were proposed for
$c=1/2$ string field theory in the temporal gauge\cite{iikmns}.
Note that the linear terms of $\delta/\delta j_0$ and $\delta/\delta j_1$
do not appear in eqs.(\ref{csd1}) and (\ref{csd2})
which would in the temporal gauge correspond to the change of the
loop length in the infinitesimal development of time.

In \cite{iikmns} the amplitude involving the loop with the
spin state of $({\cal H}(0))^2\pl$ was found to vanish when there is no other
$w(l,\hp)$ or $w(l,({\cal H}(0))^2\pl)$.
Eq.(\ref{sd3}) shows that this loop operator corresponds to
\be V_2(n)=W_2(n)-{1\over\la}W_1(n-1)+{\mu\over\la}W_0(n-1)\label{v2}.\ee
We see another operator mixing here.
In the continuum limit we have
\be y_c^n(V_2(n)-c_2(n))\sim a^5w(l,({\cal H}(0))^2\pl),\ee
since the dimension of $w(l,({\cal H}(0))^2\pl)$ is $L^5$.
We find from eqs.(\ref{sd3}) and (\ref{v2}) that
\be c_2(n)={\mu\over \la}\de{n-1}.\ee
After the subtraction of $c_2$ eq.(\ref{sd3}) leads to
\be \dd{j_2(l)}Z[j_0,j_1,j_2]|_{j_1=j_2=0}=0,\ee
where we included $j_2(l)$ in $Z$ again.
We see that we in fact have the term $l^3\delta/\delta j_2(l)$
in eq.(\ref{csd2}) since we substituted eq.(\ref{sd3}) into (\ref{sd2}).

In order to obtain the $W_3$ constraints from eqs.(\ref{csd1}) and (\ref{csd2})
we have to solve eq.(\ref{csd1}) for $\delta/\delta j_1$
and substitute it into eq.(\ref{csd2}).
Dropping $l$ from eq.(\ref{csd1}) and $l^3$ from eq.(\ref{csd2}),
we have
\be \left[\left(\dd{j_0}*+g(lj_0)\triangleleft\right)\dd{j_0}(l)
+\dd{j_1(l)}\right]Z[j_0,j_1]|_{j_1=0}+\delta(l)R_0[j_0]=0,\label{csd11}\ee
\be \left[\left(\dd{j_0}*+g(lj_0)\triangleleft\right)\dd{j_1}(l)
+16\delta^{(4)}(l)\right]Z[j_0,j_1]|_{j_1=0}
+\sum_{i=0}^2 \delta^{(i)}(l)S_i[j_0]=0.
\label{csd22}\ee
We can now solve eq.(\ref{csd11}) for $\delta Z/\delta j_1$
and substitute it into eq.(\ref{csd22}).
Following \cite{iikmns} we factor out the singular part $Z_{sin}$,
use variables of eq.(\ref{jr}) with $j$ replaced by $j_0$
where $r\equiv 1/3, 2/3 \pmod{1}$ and $r>0$,
use notations (\ref{ddd}) and (\ref{kkk})
and put $g=1$ for simplicity.
We find
\bea \lefteqn{}&&\left\{\left[\mbox{\Large :}\left(D(l)+{K(l)\over 3}
+w_s(l)\right)^3\mbox{\Large :}\right]_{\geq 0}-16t\delta^{(2)}(l)
    +3\cdot 2^{8\over 3}\delta^{(1)}(l)\left({\partial\over
      \partial j_{{2\over 3}}}
      +{2\over 9}j_{-{2\over 3}}\right)\right.\nonumber\\
&&\mbox{}
    +\delta(l)\left(3\cdot 2^{8\over 3}\left({\partial\over
      \partial j_{{5\over 3}}}
      +{5\over 9}j_{-{5\over 3}}\right)+{16\over 3}t^2\right)
    +{3\over 2}\left(D(l)-{K_q(l)\over 3}+w_s(l)
    +{2\over 3}\hat{K}\right)
\nonumber\\
&&\left.
    \left(\left[\mbox{\Large :}\left(D(l)+{K(l)\over 3}+w_s(l)\right)^2
\mbox{\Large :}+{2\over 27}l\right]_{\geq 0}
          +\delta(l)\left(2^{7\over 3}{\partial\over \partial j_{{1\over 3}}}
          +{2^{7\over 3}\over 9}j_{-{1\over 3}}\right)
\right)\right\}Z_{reg}[j_0]
\nonumber\\
&&\mbox{}
-{\sum_{i=0}^2 \delta^{(i)}(l)S_i[j_0]\over Z_{sin}[j_0]}
+\left(D(l)-{K_q(l)\over 3}+w_s(l)
    +{2\over 3}\hat{K}\right)\delta(l){R_0[j_0]\over Z_{sin}[j_0]}=0,
\label{w3}\eea
where
\be w_s(l)=2^{4\over 3}\left({l^{-{7\over 3}}\over \Gamma(-{4\over 3})}
            -{t l^{-{1\over 3}}\over 3\Gamma({2\over 3})}\right),\ee
\be \hat{K}f(l)=\int_{-l}^\infty dl'l'j_0(l')f(l+l'),\ee
\be K_q(l)=\left({l^{-{2\over 3}}\over\Gamma({1\over 3})}*\hat{K}
        {l^{-{4\over 3}}\over\Gamma(-{1\over 3})}\right)(l)
        +\left({l^{-{1\over 3}}\over\Gamma({2\over 3})}*\hat{K}
        {l^{-{5\over 3}}\over\Gamma(-{2\over 3})}\right)(l).\ee
In eq.(\ref{w3}) the $W_3$ constraints appear in the coefficient
of the $l$ expansion.
The backgrounds of $j_r$ corresponding to $w_s$ are
$-3\cdot 2^{4/3}t$ for $j_{1/3}$ and $9\cdot 2^{4/3}/7$ for $j_{7/3}$.
The terms of the $\delta$-function in eq.(\ref{w3}) lead to
\be R_0[j_0]=-2^{4\over 3}Z_{sin}[j_0]\left(
3{\partial\over \partial j_{{1\over 3}}}
           +{j_{-{1\over 3}}\over 3}\right)Z_{reg}[j_0],\ee
\be S_0[j_0]=Z_{sin}[j_0]\left(3\cdot 2^{8\over 3}\left({\partial\over
      \partial j_{{5\over 3}}}
      +{5\over 9}j_{-{5\over 3}}\right)+{16\over 3}t^2\right)Z_{reg}[j_0],
\ee
\be S_1[j_0]=3\cdot 2^{8\over 3}Z_{sin}[j_0]\left({\partial\over
             \partial j_{{2\over 3}}}+{2\over 9}j_{-{2\over 3}}\right)
            Z_{reg}[j_0],\ee
\be S_2[j_0]=-16tZ[j_0],\ee
which were not clear in \cite{iikmns}.
Here $R_0[j_0]$ and $S_i[j_0]$ correspond to $\partial/\partial J_0(1)$,
$\partial/\partial J_1(1)$
and $\partial/\partial J_1(2)$ in eqs.(\ref{sd111}) and (\ref{sd222}).
We can verify that the disk amplitudes obtained from eqs.(\ref{csd1})
and (\ref{csd2}) indeed satisfy eqs.(\ref{csd11})
and (\ref{csd22}) with $R_0[j_0]$ and $S_i[j_0]$ given above.

In this paper we have taken the continuum limit of the one and
two matrix model and obtained the continuum \sdeq s.
The resulting equation for the one matrix model reproduces
that of \cite{IK}.
The resulting equations for the two matrix model coincide with those
proposed in \cite{iikmns}.
We therefore have verified the assumption on which $c=1/2$ string
field theory in the temporal gauge was constructed.
The same technique  may be applied to the general configuration
of the spin state on the loop and to the $m-1$ matrix model to obtain
the continuum \sdeq s for $c=1-6/m(m+1)$.

\vskip 1cm

\begin{center}
{\bf Acknowledgement}
\end{center}

The auther would like to thank H. Kawai and N. Ishibashi for useful discussions
and B. Hanlon for carefully reading the manuscript.

\end{document}